\documentclass{aa}

\usepackage{graphicx}
\usepackage{txfonts}
\usepackage{natbib}
\usepackage[colorlinks=true,citecolor=blue]{hyperref}
\usepackage{xspace}

\bibpunct{(}{)}{;}{a}{}{,}

\newcommand{\MJup}{M$_{\mathrm{Jup}}$\xspace}
\newcommand{\RJup}{R$_{\mathrm{Jup}}$\xspace}

\newcommand{\teff}{T$_{e\!f\!f}$\xspace}
\newcommand{\logg}{log~\emph{g}\xspace}

\newcommand{\mic}{$\mu$m\xspace}
\newcommand{\as}{\hbox{$^{\prime\prime}$}\xspace}
\newcommand{\degre}{$^\circ$\xspace}

\begin{document}

\title{High-contrast spectroscopy of SCR~J1845-6357~B\thanks{Based on observations collected at the European Southern Observatory, Chile, ESO program 486.L-0077.}}
\subtitle{}

\author{A. Vigan\inst{1} \and M. Bonnefoy\inst{2} \and G. Chauvin\inst{3,2} \and C. Moutou\inst{4}  \and G. Montagnier\inst{5}}

\institute{Astrophysics group, School of Physics, University of Exeter, Stocker Road, Exeter EX4 4QL, United Kingdom \\
\email{arthur@astro.ex.ac.uk}
\and
Max Planck Institute for Astronomy, K\"onigstuhl 17, 69117 Heidelberg, Germany
\and
Institut de Plan\'etologie et d'Astrophysique de Grenoble, UMR 5274, CNRS, Universit\'e Joseph Fourier, Grenoble 38041, France
\and
Laboratoire d'Astrophysique de Marseille, UMR 6110, CNRS, Universit\'e de Provence, 38 rue Fr\'ed\'eric Joliot-Curie, 13388 Marseille Cedex 13, France
\and
European Southern Observatory, Casilla 19001, Santiago 19, Chile \\
}

\date{Preprint version: 1 April 2011 (it's not a joke).}

 
\abstract
{Spectral characterization of sub-stellar companions is essential to understand their composition and formation processes. However, the large contrast ratio of the brightness of each object to that of its parent star limits our ability to extract a clean spectrum, free from any significant contribution from the star. During the development of the long slit spectroscopy (LSS) mode of IRDIS, the dual-band imager and spectrograph of SPHERE for the Very Large Telescope (VLT), we proposed a data analysis method to estimate and remove the contributions of the stellar halo and speckles to the spectra.}
{This method has never been tested on real data because of the lack of instrumentation capable of combining adaptive optics (AO), coronagraphy, and LSS. Nonetheless, a similar attenuation of the star can be obtained using a particular observing configuration where the slit is positioned on the faint companion while keeping the bright primary outside.}
{Test data were acquired with this slit configuration using the AO-assisted spectrograph VLT/NACO. We obtained new $J$- and $H$-band spectra of SCR~J1845-6357~B, a T6 companion to a nearby ($3.85 \pm 0.02$~pc) M8 star. This system is a well-suited benchmark as it is relatively wide ($\sim$1.0\as) with a modest contrast ratio ($\sim$4~mag), and a previously published $JHK$ spectrum is available for reference.}
{We demonstrate that (1) our method is efficient at estimating and removing the stellar contribution, (2) it allows to properly recover the overall spectral shape of the companion spectrum, and (3) it is essential to obtain an unbiased estimation of physical parameters. We also show that the slit configuration associated with this method allows us to use long exposure times with high throughput producing high signal-to-noise ratio data. However, the signal of the companion gets over-subtracted, particularly in our $J$-band data, compelling us to use a fake companion spectrum to estimate and compensate for the loss of flux. Our spectral analysis provides an estimation of \teff~=~$1000 \pm 100$~K leading to R~=~$0.7 \pm 0.1$~\RJup, a value that closely agrees with evolutionary models for ages older than 1.5~Gyr. Finally, we report a new astrometric measurement of the position of the companion (sep = 0.817\as, PA = 227.92~deg), which has undergone a significant proper motion since the previous measurement.}
{}


\maketitle

\section{Introduction}
\label{sec:introduction}

Near-infrared emission spectra of sub-stellar companions directly imaged around pre-main sequence and main sequence stars is crucial when it comes to characterizing their atmospheric (e.g. composition and cloud properties) and physical (effective temperature \teff, surface gravity \logg, radii, mass and luminosity) properties, as well as understanding their formation and evolutionary processes. Obtaining high-quality spectra for faint companions at close angular separations from bright stars is challenging, as one needs to overcome the large contrast ratio in the brightnesses of the two objects. Low-mass companions are usually only slightly brighter than the point spread function (PSF) wings, making it necessary to develop methods to estimate and subtract at all wavelengths the stellar contribution, which can bias the spectral analysis. This will be even truer for upcoming high-contrast coronagraphic imagers such as SPHERE \citep{beuzit2008} for the Very Large Telescope (VLT) or GPI \citep{macintosh2008} for Gemini, where the images and spectra of exoplanets will be strongly limited by the speckle noise associated with the star \citep{soummer2007}. 

In the context of the development of the differential spectro-imager SPHERE/IRDIS \citep{dohlen2008}, we proposed in \citet{vigan2008} a new data analysis method to estimate and subtract the stellar halo and speckles in high-contrast long slit spectroscopy (LSS) data. The method was designed to attenuate the speckle noise and extract a clean companion spectrum, free from any significant stellar contribution. It has proven to be very efficient in realistic simulated data for SPHERE/IRDIS, but has never been tested in real data. To demonstrate the method, we proposed using the VLT/NACO AO-fed spectrograph \citep{lenzen2003,rousset2003} in a particular configuration where the slit is positioned on a faint sub-stellar companion and the bright primary is kept outside of the slit to avoid saturation. Although not strictly identical to what will be available in SPHERE/IRDIS, this configuration improves the dynamical range compared to the more standard configurations where the bright primary and the sub-stellar companion are kept within the slit \citep[e.g.][]{mohanty2007,lafreniere2008,janson2010}.

\object{SCR~J1845-6357} -- hereafter SCR~1845 -- is a close ($3.85 \pm 0.02$~pc; \citealt{henry2006}) M8.5 star \citep{henry2004} discovered in the context of the SuperCOSMOS survey \citep{hambly2001}. \citet{biller2006} discovered a bound companion orbiting at $\sim$4.5~AU from the primary using VLT/NACO AO-imaging, making it the first system with a tight brown-dwarf companion to a main-sequence star. They estimated SCR~1845~B to be a T4.5-T6.5 brown dwarf based on its colors in narrow-band filters centered inside and outside of the 1.6~\mic CH$_{4}$ absorption band (``spectral differential imaging'' mode). The T6 spectral type was later confirmed using NACO spectroscopic observations \citep{kasper2007}, which allowed them to constrain \teff and \logg to be $\sim$950~K and 5.1~dex respectively, assuming solar metallicity. They also estimated an age ranging from 1.8~Gyr to 3.1~Gyr for this system, which translates into a mass of 41.0--51.8~\MJup for the companion using the evolutionary models of \citet{burrows1997}. SCR~1845~B is then very similar in terms of distance, age, and spectral type to the well-known T1/T6 dwarf binary orbiting the K4.5V star $\epsilon$~Indi~A \citep{mccaughrean2004,king2010}.

Systems such as SCR~1845 and $\epsilon$~Indi~Ba/Bb are extremely appealing for the calibration of evolutionary models: their proximity to the Sun and their small orbital radius make them excellent candidates for an independent, dynamical estimation of the system mass on relatively short timescales. It is all the more important that evolutionary tracks are not well-calibrated at low masses and young ages \citep[e.g.][]{konopacky2010}. In the present work, we provide a new astrometric measurement of the position of SCR~1845~B, which has a large proper motion with respect to previous measurements, and we obtain new VLT/NACO $J$- and $H$-band spectra to test our data analysis method. We start by giving a brief description of the observations and basic data reduction steps in Sect.~\ref{sec:observations_data_reduction}, before describing in detail the results, performance, and problems of our data analysis in Sect.~\ref{sec:spectral_deconvolution}. In Sect.~\ref{sec:new_spectral_analysis}, we discuss our new spectral extraction of SCR~1845~B, and finally in Sect.~\ref{sec:discussion} we discuss the impact of suppressing the stellar contribution and compare our data analysis method to previously published methods in the context of future high-contrast spectrographs.

\section{Observations and data reduction}
\label{sec:observations_data_reduction}

\begin{figure}
  \centering
  \includegraphics[width=0.5\textwidth]{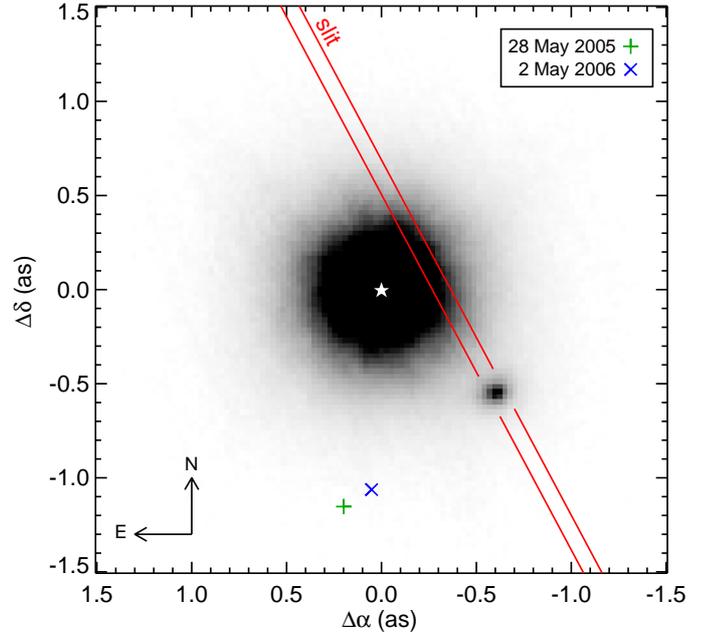}
  \caption{VLT/NACO image of SCR~1845~AB obtained in $H$-band with the S27 camera on 20 October 2010. North is up and east is left. The center of the primary is symbolized by a white star symbol. The relative positions of SCR~1845~B reported by \citet{biller2006} and \citet{kasper2007} are overlaid on the image, respectively, as a green plus sign and blue cross. The peculiar orientation adopted for the slit in our work is also overlaid on the image.}
  \label{fig:relative_astrometry}
\end{figure}

\begin{table}
  \caption{SCR~1845~B astrometry}
  \label{tab:relative_astrometry}
  \centering
  \begin{tabular}{lccc}
  \hline\hline
  \multicolumn{1}{c}{Date} & Separation & PA           & Ref. \\
               & (as)              & (deg)             &      \\
  \hline
  ~~1 May 2005 & $1.176 \pm 0.001$ & $170.22 \pm 0.08$ & (1)  \\
  28 May 2005  & $1.170 \pm 0.003$ & $170.20 \pm 0.13$ & (2)  \\
  ~~2 May 2006 & $1.064 \pm 0.004$ & $177.20 \pm 0.06$ & (3)  \\
  20 Oct. 2010 & $0.817 \pm 0.003$ & $227.92 \pm 0.46$ &      \\
  \hline
  \end{tabular}
\begin{flushleft} References: (1) \citet{montagnier2006}; (2) \citet{biller2006}; (3) \citet{kasper2007}.\end{flushleft}
\end{table}

\subsection{Imaging}
\label{sec:imaging}

Data were obtained on 20 October 2010 with the instrument VLT/NACO in service mode (ESO program 486.L-0077). The infrared wavefront sensor with $14 \times 14$ lenslets array was used together with a dichroic transmitting 90\% of the light in K band to the NAOS wavefront sensor and 90\% of the light in either J or H band to CONICA. We took advantage of the median-quality atmospheric conditions in the $J$- (seeing of $\sim$1.1\as, airmass of $\sim$1.5) and $H$- (seeing of $\sim$0.8\as, airmass of $\sim$1.4) bands, providing a good AO correction.

A series of 5$\times$5~sec exposures were recorded with the S27 camera and the $H$-band near-infrared filter, providing a spatial scale of 27.15~mas/pixel. These frames were divided by a lamp flat-field exposure and corrected for the effects of bad pixels using customized IDL routines. The cleaned frames were then median-combined to produce a final image (see Fig.~\ref{fig:relative_astrometry}). Since we did not foresee performing precise photometry, no dithering procedure was applied during the observations and no sky frame was subtracted. Precise relative astrometry was performed by fitting two-dimensional (2D) Gaussian functions to both components of the system. Table~\ref{tab:relative_astrometry} compares our relative astrometry to previously reported values. Given the proximity of SCR~1845 to the Sun and the short angular separation of the system, the orbital motion has been significant in the past five years.

\subsection{Spectroscopy}
\label{sec:spectroscopy}

\begin{figure*}
  \centering
  \includegraphics[width=1.0\textwidth]{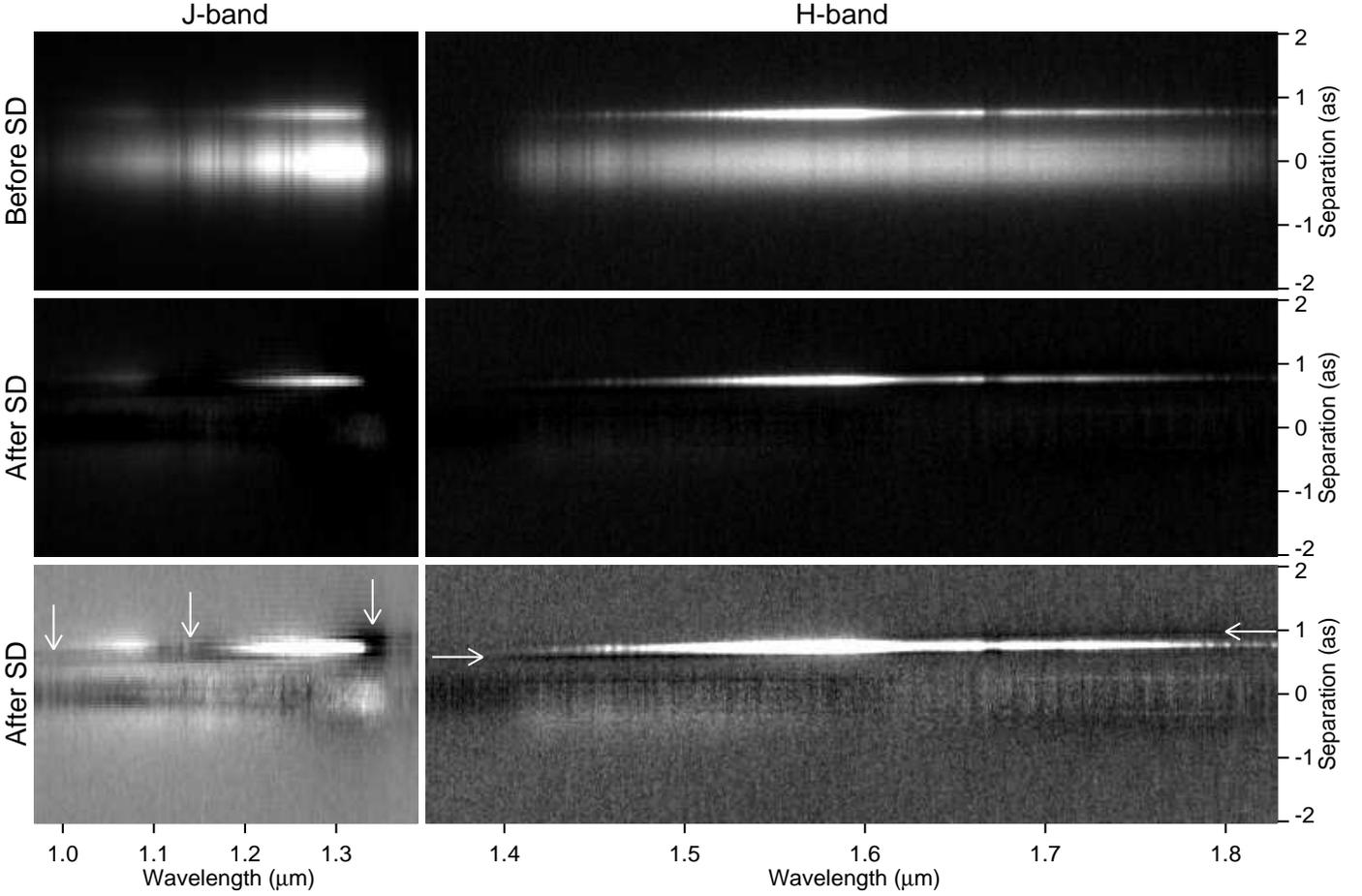}
  \caption{Comparison of the $J$- (left) and $H$-band (right) spectra before and after the spectral deconvolution data analysis used to remove the contribution of the star. The top panels display the spectra with no star subtraction, the middle panels the spectra after data analysis at the same display scale as the top panels, and the bottom panels the spectra after data analysis with an appropriate display scale to identify the areas of over-subtraction (darker areas). The areas of strong over-subtaction are identified using small white arrows in the bottom panel. \textbf{The display scales are different in the $J$- and $H$-bands.}}
  \label{fig:reduced_spectra}
\end{figure*}

NACO was used in LSS mode to obtain $R$~=~400--500 spectra in $J$- and $H$-band with the 86~mas slit. The spectroscopic modes used were S54-4-SJ for $J$-band, providing $\sim$2~nm/pixel spectral dispersion over the range 0.91--1.40~\mic and 54.60~mas/pixel spatial resolution, and S27-4-SH for $H$-band, providing $\sim$1~nm/pixel spectral dispersion over the range 1.37--1.84~\mic and 27.15~mas/pixel spatial resolution. The slit was aligned in a very particular way to ensure that the bright core of the PSF of the primary did not enter the slit (see Fig.~\ref{fig:relative_astrometry}), allowing a remarkable increase in sensitivity to the companion signal, while keeping signal from the star, which will be estimated and subtracted during the data analysis. A standard nodding sequence of 10\as was used to obtain eight exposures of 70~sec (DIT$\times$NDIT = 10$\times$7), resulting in a 560~sec total exposure time, in both the $J$- and $H$-bands. A telluric calibrator  (HIP~89729, spectral type O5.5III) was observed just after the object at a similar airmass. A spectrum of the NACO internal argon arc lamp was recorded for wavelength calibration.

The spectroscopic data was reduced using \texttt{REDSPEC}\footnote{\url{http://www2.keck.hawaii.edu/inst/nirspec/redspec.html}}, the data reduction software originally designed for the high-resolution spectrograph Keck/NIRSPEC. The software was used to perform the following data reduction steps: division by flat-field, correction of bad pixels, correction of the spectrum distortion, and wavelength calibration. The difference between pairs of corrected frames was obtained to eliminate the sky contribution. The frames were recentered on the same spatial position using the signal of the companion as a reference, and finally median-combined to create the final reduced spectrum. The resulting spectra in $J$- and $H$-band are visible in the top panels of Fig.~\ref{fig:reduced_spectra}.

\section{Spectral deconvolution}
\label{sec:spectral_deconvolution}

\subsection{Context}
\label{sec:context}

\begin{figure*}
  \centering
  \includegraphics[width=1.0\textwidth]{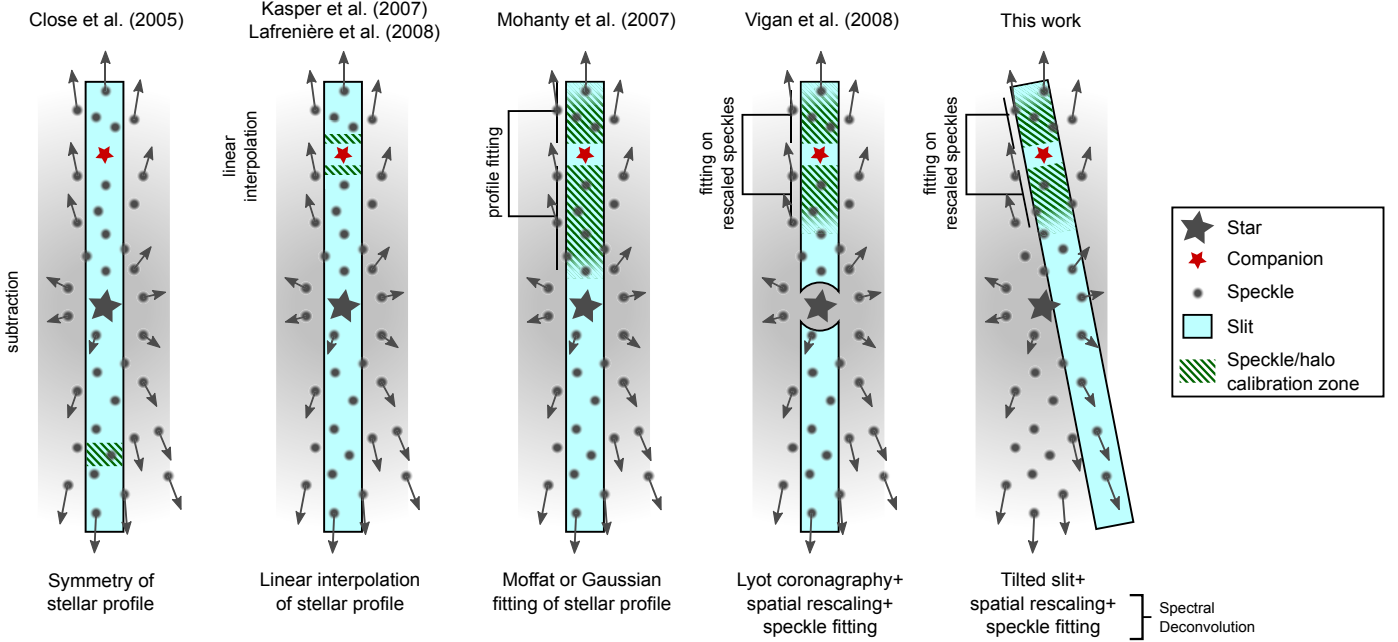}
  \caption{Illustration of the various data analysis methods that have been proposed and used to estimate and subtract the stellar halo and speckles at the position of a faint companion. The star, halo, and speckles are represented in dark grey, respectively, with a five-pointed star, a wide radial gradient, and small grey circles; the companion is represented by a smaller red five-pointed star; the portion of starlight used to calibrate the halo/speckles at the position of the companion is represented by a hashed green area. The chromatic motion of the halo and speckle pattern is represented by dark grey arrows starting from the speckles and pointing outwards (i.e. motion for increasing wavelength). A short explanation of the methods is given beneath each illustration. In the \citet{vigan2008} configuration, the slit is coupled with a Lyot coronagraph, represented by a circular focal-plane mask at the center of the slit.}
  \label{fig:data_analysis_methods}
\end{figure*}

In high-contrast imaging and spectroscopy, the signal of the companion is usually close to or below the brightness level of the stellar halo and speckles, requiring the use of specific data analysis methods to alleviate the contribution of the star, which otherwise contaminates the spectral extraction of the companion. This is extremely important (i) to accurately retrieve the shape of the companion continuum and (ii) to isolate the spectral features of the companion. In Fig.~\ref{fig:reduced_spectra} (top panels), we see that the companion signal (thin straight line) is overlaid on the stellar halo and speckles. The signal of the companion is then biased by the star signal, so the goal of the data analysis step will be to obtain a spectrum of the companion that is free of any significant contribution from the star. 

The differentiation of point source spectra from a constant or varying background is a problem encountered in very different contexts with LSS, and several solutions have been proposed. An iterative approach was proposed by \citet{lucy2003} and successfully used to extract clean supernova spectra \citep{blondin2005}. This method is based on a two-channel algorithm that restores a PSF-like component in a 2D image and an underlying extended background separately. However, it requires precise knowledge of the slit-spread function (SSF) to restore the point-source channel, and a misevaluation of the width of the spatial resolution kernel necessary for the point-source/extended background discrimination will affect the restoration \citep[see][Sect.~3]{blondin2005}. This method has so far never been used to extract faint sub-stellar companion spectra.

More specifically to sub-stellar companions, \citet{close2005} used the symmetry of the stellar profile to extract the spectrum of AB~Dor~C: they subtracted the contribution of AB~Dor~A at the position of the companion using a spectrum acquired with the telescope derotator rotated by 180\degre. For the spectrum of 2M~1207~b, \citet{mohanty2007} modeled the stellar halo in each spectral channel using a sum of three Gaussians plus a second order polynomial. For the spectrum of 1RXS~1609~b, \citet{lafreniere2008} simply interpolated the star contribution with a straight line underneath the companion signal. This assumption, although quite simplistic, is well-justified by the wide separation of the system (2.22\as). The same approach was used by \citet{kasper2007} for the spectrum of SCR~1845~B. Figure~\ref{fig:data_analysis_methods} presents an illustration of these different data analysis methods with the relevant information: position of the star, halo and speckles, position of the companion, orientation of the slit, chromatic motion of the speckles, and portion of starlight used to calibrate the halo and speckles.

At higher contrasts, we developed a method for the SPHERE/IRDIS LSS mode \citep{vigan2008} based on the ``spectral deconvolution'' -- hereafter SD -- originally proposed by \citet{sparks2002}. It is based on the linear dependence of the position and size of all diffraction-related patterns (stellar halo, speckles, Airy rings) on wavelength, when the position of an object in the field, such as a sub-stellar companion, is fixed. The main difference from the present work is that IRDIS will combine LSS with coronagraphy to be able to characterize faint companions around nearby bright stars. In the case of IRDIS, the diffraction patterns will then move away radially from the star inside the slit with increasing wavelength. 

In the present work, we slightly tilted the slit to decenter the star (see Fig.~\ref{fig:data_analysis_methods}), allowing us to use long exposure times without saturating the star signal. In this configuration, the speckles still move away radially from the star, but because of the tilted slit they do not necessarily remain inside the slit over the full wavelength range. This purely gometrical effect is illustrated in Fig.~\ref{fig:data_analysis_methods} (rightmost illustration): the closer a speckle is to the star, the more orthogonal to the slit its chromatic motion is. The result is that speckles at small angular separations will either enter or exit the slit, creating a variation in luminosity in the final spectrum that cannot be calibrated, while speckles at larger separations (e.g. close to the companion) will remain in the slit over a wider range of wavelengths, allowing an a posteriori calibration of their luminosity variation with wavelength. Another way of stating the problem is to say that because of the perspective effect, the slit is much wider at larger angular separations from the direction of the star. In our data, the separation of the companion is large enough (0.817\as) to ensure that this effect does not prevent us from applying the SD. Moreover, at this separation the data is mostly dominated by the star halo, which shows smoother variations than bright speckles very close to the star. In the following sections, we focus on validating our data analysis method assuming that this effect does not seriously hamper the analysis at the position of the companion.

\subsection{Application to the data}
\label{sec:application_to_data}

\begin{figure}
  \centering
  \includegraphics[width=0.5\textwidth]{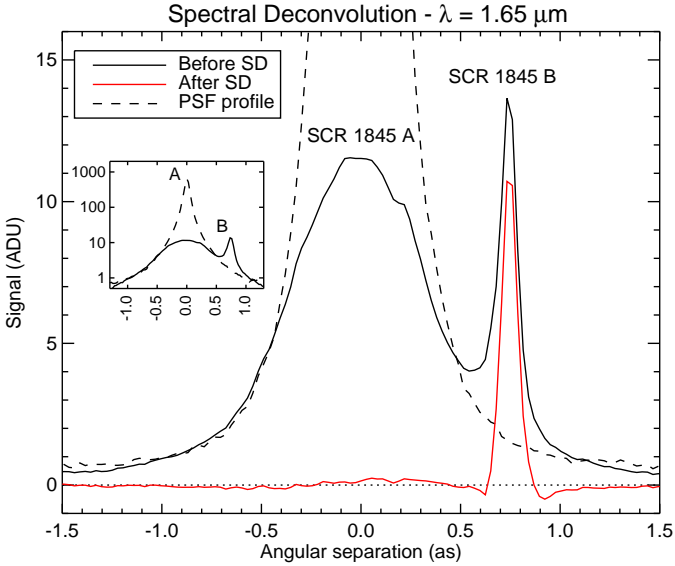}
  \caption{Comparison of cuts along the spatial dimension of the spectra before (plain, black) and after (plain, red) SD at $\lambda = 1.65$~\mic. Since the star was kept outside the slit to prevent saturation, the profile of the primary before SD does not show the true peak of the primary PSF, but instead the level of the halo brightness with the slit decentered by $\sim$0.2\as. The PSF profile from the star obtained from Fig.~\ref{fig:relative_astrometry} is overplotted (dashed, black) to give an estimation of the true contrast (4.17~mag in $H$-band) between the companion and the primary. For clarity, the main plot has been cut at 16~ADU, but the sub-plot shows the full profiles on a log scale with the same units in x and y.}
  \label{fig:spectrum_cut_SH}
\end{figure}

The SD was applied separately to the $J$- and $H$-band spectra obtained at the end of the reduction steps detailed in Sect.~\ref{sec:observations_data_reduction}. The result is displayed in the middle panels of Fig.~\ref{fig:reduced_spectra}. Compared to the spectra before SD, the contribution of the star has been significantly attenuated, although some residuals are still visible at the position of the star. This is particularly visible when using an appropriate display scale (Fig.~\ref{fig:reduced_spectra} bottom panels). Two effects are clearly visible:

\begin{itemize}
\item The non-zero residuals at the position of the star signal because, as previously explained, the diffraction-related patterns move through the slit. This effect makes it difficult to precisely evaluate their contribution close to the star, where their chromatic motion is almost orthogonal to the slit and they are brightest. This effect is significant within $\pm$0.5--0.6\as of the position of the star. \\

\item the over-subtraction visible on either side of the companion spectrum because the estimation of the stellar contribution at the position of the companion in the rescaled spectrum is biased by the companion signal. To mitigate this effect (but not cancel it), the companion signal is masked in the fitting process with a 5$\lambda/D$ aperture.
\end{itemize}

\noindent The residuals at the position of the star are not problematic because we attempt to extract the spectrum of the faint companion, so we do not study its effect in any more detail. While in H band the over-subtraction appears as thin areas either below or above the companion spectrum (see white arrows in Fig.~\ref{fig:reduced_spectra}), it appears much stronger in $J$-band, especially at the center and the red-end of the spectrum, making it impossible to recover a spectrum over the full band (see Sect.~\ref{sec:validation_method}). The over-subtraction is analyzed in more detail in the next section. Figure~\ref{fig:spectrum_cut_SH} illustrates the effect of the data analysis with a cut of the spectrum before and after SD in $H$-band at $\lambda = 1.65$~\mic. Since the star was kept outside the slit, the profile before SD actually shows the level of the star halo when the slit was decentered by $\sim$0.2\as. The level of the PSF profile obtained from Fig.~\ref{fig:relative_astrometry} is overplotted to show the true brightness contrast difference of the companion with respect to the primary (4.17~mag in $H$-band). The effect of the SD is clearly visible, as the star contribution was significantly reduced, leaving mostly residuals with an average close to zero. On either side of the companion signal, the over-subtraction appears as negative areas of 0.1\as--0.2\as width. Apart from these areas, the signal of the companion is accurately recovered, demonstrating that our data analysis method is successful at estimating and subtracting the stellar contribution. 

Finally, the one-dimensional spectrum of the companion is obtained by summing the signal in each spectral channel over a four pixel-wide aperture centered on the position of the companion. The same procedure is applied to the telluric calibrator observed after SCR~1845. The companion spectrum is then divided by the calibrator spectrum and multiplied by a black body at 44\,500~K (\teff of the telluric calibrator) to obtain the final spectrum that is compared to other methods and used for the spectral analysis.

\subsection{Impact of the over-subtraction}
\label{sec:impact_over_subtraction}

\begin{figure}
  \centering
  \includegraphics[width=0.5\textwidth]{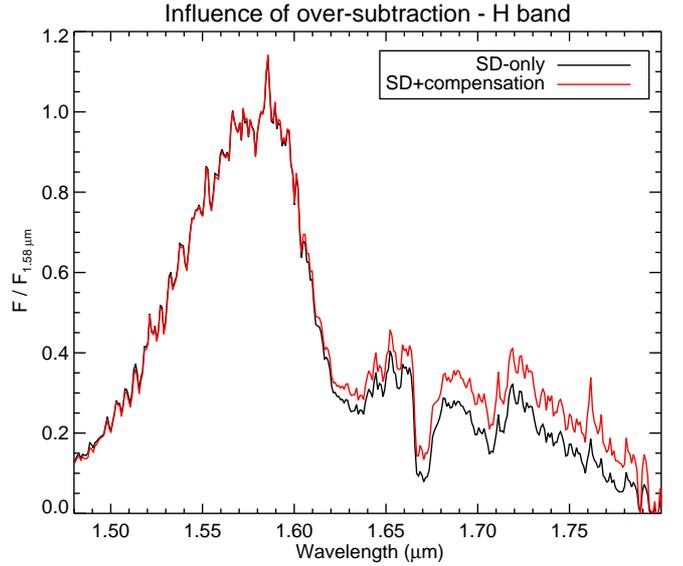}
  \caption{Comparison of the SD-only and SD+compensation spectra in $H$-band. The SD+compensation spectrum corresponds to the SD-only spectrum corrected for the over-subtraction by introducing a fake companion into the data.}
  \label{fig:effect_flux_loss}
\end{figure}

The SD data analysis method relies on two assumptions: (1) the data is sufficiently well sampled spatially to ensure that it can be properly rescaled by interpolation without introducing artifacts into the data, and (2) the star ``model spectrum'' can be properly estimated and then fitted in each spatial channel to remove the star contribution. If these two conditions are not met, the estimation of the stellar halo and speckles will be biased, and the companion spectrum will not be recovered properly. 

In our data, the condition (1) is far from being met in $J$-band, where the $\sim$54~mas/pixel spatial sampling is 4.2 times larger than the Shannon sampling (12.9~mas/pixel). The impact of this under-sampling is visible in Fig.~\ref{fig:reduced_spectra} (bottom panels) as horizontal stripes appearing on both sides of the companion signal in $J$-band owing to the interpolation. With a 27~mas/pixel scale, the $H$-band was undersampled by a factor of only 1.4, and no clear indication of under-sampling was visible in the data analysis. Although under-sampling is problematic from a signal-processing perspective, we verified that it is not the main limitation here by binning the $H$-band data to the same spatial sampling as the $J$-band data. The resulting spectrum was similar to the one without binning within 5\%.

The condition (2) can be met by masking the companion signal in the data analysis (as detailed in \citealt{vigan2008}). At a given angular separation, the size of the mask is limited by the value of the \emph{bifurcation point} \citep{thatte2007}. The bifurcation point represents the smallest angular separation at which the SD can be applied, which depends on the wavelength range of the data and the spatial extent of the companion signal. It measures the separation at which the spatial rescaling of the data will move the companion signal by at least its spatial extent, allowing us to fit the stellar speckles and halo on either side of the companion signal (see \citealt[][Fig.~2]{vigan2008} for an example of rescaled data). For companions at separations below the bifurcation point, the spatial rescaling will not move the companion signal enough to fit the speckles, and the SD will not work.

If the wavelength range and the separation of the companion are fixed, as in our observations, the bifurcation relations \citep[][Eq.~2 and 3]{thatte2007} can be used to calculate the largest size of the mask that can be applied to the companion signal to hide it while still being able to use SD. If the size of the mask is larger than the maximum set by the bifurcation point, there will not be enough pixels to estimate the star contribution and the final result will be biased. Although the mask size was chosen to be an adequate size, the effect of the companion signal still influences the data analysis outside the masked region. A possible explanation is that since NAOS is not a high-order AO system, the PSF is not fully diffraction-limited, so it has a full width at half-maximum (FWHM) larger than $\lambda/D$. The FWHM was estimated to be $\sim$100~mas at 1.65~\mic in our imaging data, while the slit has a width of only 86~mas, meaning that the companion PSF is larger than the slit. Diffraction by the edge of the slit then broadens the signal on the detector. The impact is stronger in $J$-band because of the poorer seeing, higher airmass, and smaller Strehl ratio inducing a larger broadening of the PSF.

To cope with the over-subtraction, a fake companion spectrum is introduced into the data at a position symmetric to that of the real companion with respect to the star. To be as accurate as possible, the level of signal of the fake spectrum in each spectral channel is chosen to match the level of the true companion in the same channel (including a stellar contribution). After SD, the fake spectrum is extracted using exactly the same procedure as described in Sect.~\ref{sec:application_to_data}, and the efficiency (i.e. the amount of conserved flux) is obtained by dividing the input fake spectrum by the extracted spectrum. Finally, the real companion spectrum is corrected for the over-subtraction by dividing it by the efficiency. The spectrum corrected for the over-subtraction is hereafter referred to as the SD+compensation spectrum. The spectrum that is not corrected will be referred as the SD-only spectrum. A comparison of both spectra in the $H$-band is plotted in Fig.~\ref{fig:effect_flux_loss}. The main changes occur in the spectral slope above 1.6~\mic, where the variation is limited to a few percent. As we demonstrate in the next section, the compensation of the flux loss is necessary to match the spectral extraction obtained with other methods. In conclusion, we note that our method for compensating for the flux loss is valid here because of the modest contrast between the companion and the primary, which allows an accurate estimation of the companion signal. For companions much fainter than the star, alternative solutions will need to be developed to compensate for the over-subtraction if this occurs.

\subsection{Validation of the method}
\label{sec:validation_method}

\begin{figure*}
  \centering
  \includegraphics[width=1.0\textwidth]{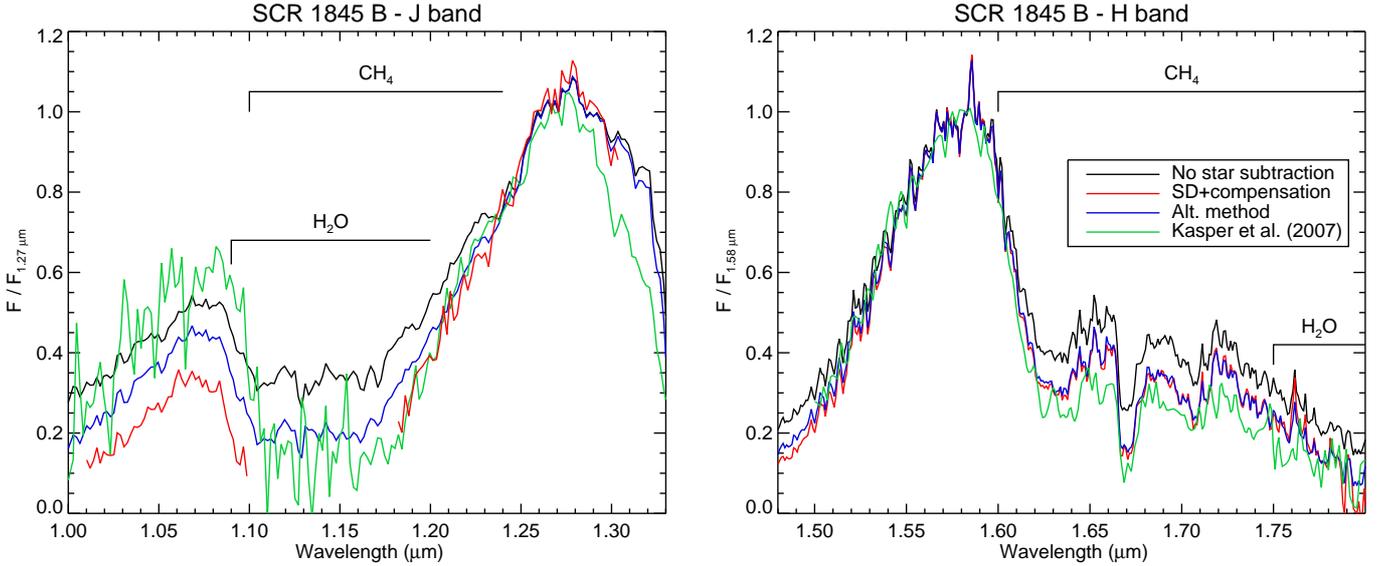}
  \caption{Comparison of the $J$-band (left) and $H$-band (right) spectra of SCR~1845~B obtained without star subtraction (black), with SD+compensation (red), and with the alternative data analysis method (blue). The spectrum  published by \citet{kasper2007} is also plotted (green).}
  \label{fig:extracted_spectra}
\end{figure*}

\begin{table}
  \caption{Value of $\epsilon_{\mathrm{rel}}$ when comparing the SD+compensation spectrum to other spectra.}
  \label{tab:comparison_spectra}
  \centering
  \begin{tabular}{cr@{.}lr@{.}l}
  \hline\hline
  Comparison spectrum & \multicolumn{2}{c}{$J$-band} & \multicolumn{2}{c}{$H$-band} \\ 
  \hline
  No star subtraction & 28&7\% & 17&3\% \\
  Alternative method  & 15&7\% &  3&9\% \\
  \citet{kasper2007}  & 35&2\% & 14&9\% \\
  \hline
  \end{tabular}
\end{table}

To validate the SD, we need to compare the extracted spectrum to one obtained after applying an independent method to the same data. For that, we use an alternative method based on an iterative process: first the stellar halo is estimated and subtracted from each spectral channel using a Moffat profile, which closely reproduces the shape of the stellar profile; the position of the companion is then measured precisely in each channel, and its signal is duplicated by symmetry around its center; the duplicated spectrum of the companion is subtracted from the original spectrum to obtain a stellar spectrum that is free of the companion contribution; and finally the stellar halo is estimated for this new spectrum using a Moffat profile in each channel, and then subtracted from the original spectrum. This method is analogous to that of \citet{mohanty2007} presented in Sect.~\ref{sec:context}. As for the SD, this method provides a 2D spectrum whose stellar contribution has been estimated and removed, but as we see in Sect.~\ref{sec:discussion} it will be of limited use for very high-contrast in observations made by future spectro-imagers.

The comparison between the different spectra for the $J$- and $H$-bands is plotted in Figure~\ref{fig:extracted_spectra}. For reference, the spectrum of the companion for which the stellar contribution was not subtracted is also plotted. To quantify the match between the SD+compensation spectrum and other spectra, we used a relative spectro-photometric error defined by

\begin{equation}
  \label{eq:relative_error}
  \epsilon_{\mathrm{rel}} = \sqrt{N_{\lambda} \sum_{p=1}^{N_{\lambda}} \left( \frac{I_{p}}{\sum_{k=1}^{N_{\lambda}} I_{k}^{\mathrm{SD}}} - \frac{I_{p}^{\mathrm{SD}}}{\sum_{k=1}^{N_{\lambda}} I_{k}^{\mathrm{SD}}} \right)^2 },
\end{equation}

\noindent where $I_{p}^{\mathrm{SD}}$ and $I_{p}$ are respectively the intensities at wavelength $p$ of the SD+compensation spectrum and the spectrum to which we compare it, and $N_{\lambda}$ the number of wavelengths. Equation~\ref{eq:relative_error} measures the relative difference between the spectral energy distributions of the two spectra, up to a scaling constant, making the measurement independent of the scaling of both spectra. Table~\ref{tab:comparison_spectra} gives the values of $\epsilon_{\mathrm{rel}}$ when comparing the SD+compensation spectrum to other spectra. We also used a classical $\chi^2$ coefficient to compare the spectra together and found that the trends were identical to the $\epsilon_{\mathrm{rel}}$ factor, so the values have not been reported in Table~\ref{tab:comparison_spectra}.

The agreement between the SD+compensation and the alternative method is visually excellent, particularly in $H$-band where the spectral slope and the molecular and atomic features are identical between the two methods. This is confirmed by the value of $\epsilon_{\mathrm{rel}}$, which is less than 4\%. In $J$-band, the strong over-subtraction induced by the SD does not allow us to recover a spectrum over the full spectral range. The large over-subtraction between 1.10~\mic and 1.18~\mic causes the signal to drop to negative unphysical values. In the properly recovered parts of the spectrum, the agreement is good above 1.2~\mic, but below 1.1~\mic the over-subtraction is not completely compensated for. The mismatch between the spectra is confirmed by the value of $\epsilon_{\mathrm{rel}}$ being four times higher than in the $H$-band, which makes us conclude that the $J$-band SD+compensation spectrum is not fully reliable. 

We can then conclude that for data properly sampled and taken in good observing conditions, as in the case of our $H$-band data, the SD allows us to attenuate the stellar contribution and recover the overall spectral shape and features of the companion. In particular, we note that the spectrum extracted without subtracting the star \emph{does not match} either the SD+compensation or the alternative method spectra: the overall spectral slope is different in both the $J$- and $H$-bands, and the depth of some strong spectral features in $H$-band (e.g. the absorption at 1.67~\mic) is slightly more important after the stellar contribution has been removed. This is confirmed in $H$-band by a value for $\epsilon_{\mathrm{rel}}$ of 17.3\%, which denotes a large discrepancy between the SD+compensation spectrum and the spectrum without stellar subtraction. The use of SD on our data analysis method is then necessary to avoid any mis-estimation of the shape of the continuum.

\subsection{Comparison to other data}
\label{sec:comparison_other_data}

The spectrum of SCR~1845~B published by \citet{kasper2007} is also plotted for comparison in Fig.~\ref{fig:extracted_spectra}. Its difference from the SD spectrum is clearly evident, particularly in $J$-band where the H$_{2}$O absorption between 1.1~\mic and 1.2~\mic is much weaker in our data, leading to a value of 35.2\% for $\epsilon_{\mathrm{rel}}$. In $H$-band, the visual agreement is much better for both the overall spectral slope and the depth of the features, but the value of $\epsilon_{\mathrm{rel}}$ is still large (14.9\%). This discrepancy, mostly due to a small difference of slope above 1.6~\mic, is compatible with slit losses, a well-known issue affecting any AO-assisted long-slit spectrographs \citep{goto2003}. The effect of slit losses generally translates into the data as variations in the spectral slope of a few percent when observing conditions change. Given that \citet{kasper2007} data was taken at a different date, with a different instrumental setup, we can then conclude that the agreement with our data is reasonable in $H$-band and does not question the validity of the SD method. In $J$-band, it is difficult to draw any conclusion as there is a large discrepancy ($\epsilon_{\mathrm{rel}} = 15.7\%$) between both our methods. 

Finally, we note that the data of \citet{kasper2007} have a much lower signal-to-noise ratio (S/N) than our data. They completed a 900~s total exposure time on target, but had to use a dichroic transmitting only 10\% of the incoming light to the science detector to avoid saturation of the primary on the slit, resulting in an effective exposure time of only 90~s. In contrast, our observations totaled 560~s of exposure time with a dichroic transmitting 90\% of the incoming light to the science camera, resulting in a much longer effective exposure time of 504~s. While this explains the noisier part of their spectrum below 1.2~\mic, it cannot fully account for the differences between our SD and the alternative method spectra, which are probably due to the differences in observing conditions. However, it is interesting to point out that our observing strategy allows us to use longer exposure times at higher throughput, resulting in higher overall S/N.

\section{New spectral analysis of SCR~1845~B}
\label{sec:new_spectral_analysis}

The significant proper motion of the companion prompted us to use the Levenberg-Marquardt algorithm \citep{press1992} to search for the orbital solution providing the minimal $\chi^2$, hence best-fit to our astrometric measurements. However, we did not find any valid orbital solutions. Although our astrometric sampling remains sparse, additional monitoring will be extremely important to confirm that this system is indeed physically bound. In the absence of results from astrometry, we then focus the current section on the results obtained from the spectral analysis of our new spectrum of SCR~1845~B. 

\subsection{Atmospheric parameters}
\label{sec:atmospheric_parameters}

\begin{figure*}
  \centering
  \includegraphics[width=1.0\textwidth]{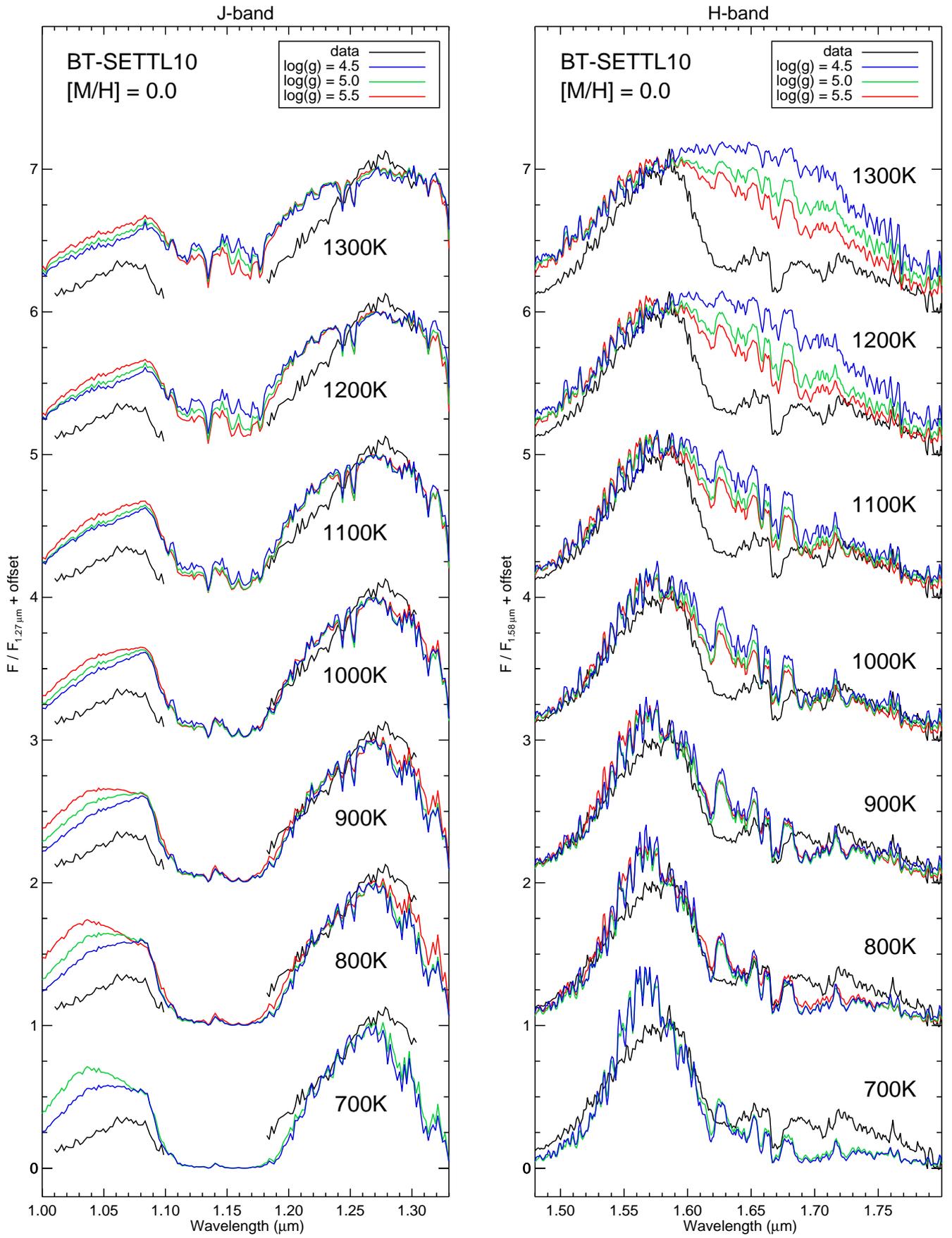}
  \caption{Comparison of the SCR~1845~B $J$- and $H$-band spectrum to the synthetic spectra of the BT-SETTL10 library for \teff from 700~K to 1300~K, \logg from 4.5 to 5.0, and [M/H]~=~0~dex. The model at \teff~=~700~K and \logg~=~5.5~dex is currently unavailable from the library.}
  \label{fig:compare_spectrum}
\end{figure*}

\begin{table}
  \caption{Spectral analysis of our different spectra of SCR~1845~B using the BT-SETTL10 library.}
  \label{tab:spectral_analysis}
  \centering
  \begin{tabular}{lccccc}
  \hline\hline
  \multicolumn{1}{c}{Spectrum} & Method LS$^{\mathrm{a}}$ & \multicolumn{2}{c}{$J$-band}  & \multicolumn{2}{c}{$H$-band}  \\
  \hline
                     &                  &	\teff 	&     \logg     &      \teff    &     \logg     \\
  \hline
  SD+compensation    &	Classic		&	 900	&	4.0	&	1100	&	4.5	\\
  SD+compensation    &	Weighted	&	1000	&	4.0	&	 900	&	5.5	\\
  \hline
  SD only	     &	Classic		&	 600	&	3.5	&	1000	&	5.0	\\
  SD only	     &	Weighted	&	 600	&	3.5	&	 800	&	5.5	\\
  \hline
  No star subtraction   &	Classic		&	1200	&	4.0	&	1200	&	5.0	\\
  No star subtraction   &	Weighted	&	1200	&	4.0	&	1200	&	5.0	\\
  \hline		
  Alt. method	     &	Classic		&	 900	&	4.0	&	1100	&	4.5	\\
  Alt. method	     &	Weighted	&	1100	&	4.0	&	1100	&	5.0	\\
  \hline		
  \citet{kasper2007} &	Classic		&	1000	&	4.5	&	1100	&	5.0	\\
  \citet{kasper2007} &	Weighted	&	 900	&	4.5	&	 900	&	5.5	\\
  \hline	
  \end{tabular}
\begin{list}{}{}
\item[$^{\mathrm{a}}$] The least squares minimization is either ``classic'' or ``weighted'' as described in \citet{mohanty2007}.
\end{list}
\end{table}

\begin{figure}
  \centering
  \includegraphics[width=0.5\textwidth]{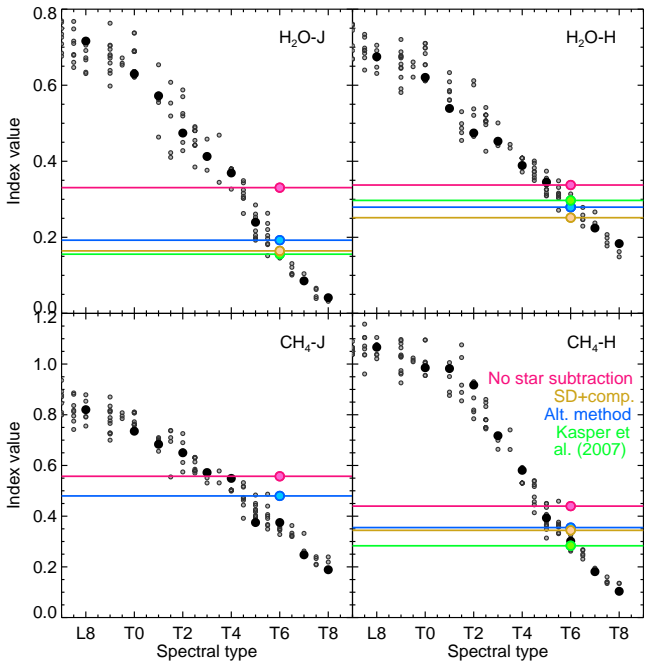}
  \caption{H$_{2}$O and CH$_{4}$ spectral indices defined by \citet{burgasser2006} for the classification of T dwarfs calculated in the spectra available for SCR~1845~B. For comparison, the grey dots represent the spectral index values for T dwarfs from the SpecX library, and the black dots the values for the reference objects from \citet[][Table 4]{burgasser2006} with available SpecX spectrum. The spectral coverage of the \citet{kasper2007} spectrum does not allow us to measure the CH$_{4}$-$J$ index. The values of all the spectral indices are summarized in Table~\ref{tab:spectral_indexes}.}
  \label{fig:spectral_indexes}
\end{figure}

\begin{table}
  \caption{Spectral index values.}
  \label{tab:spectral_indexes}
  \centering
  \begin{tabular}{lcccc}
  \hline\hline
  \multicolumn{1}{c}{Spectrum} & H$_2$O-$J$ & H$_2$O-$H$ & CH$_4$-$J$ & CH$_4$-$H$ \\
  \hline
  SD+compensation              & 0.164      & 0.252      & 0.866      & 0.344      \\
  No star subtraction          & 0.331      & 0.338      & 0.557      & 0.440      \\
  Alt. method	               & 0.192      & 0.279      & 0.480      & 0.355      \\
  \citet{kasper2007}           & 0.156      & 0.297      & \ldots     & 0.283      \\
  \hline	
  \end{tabular}
\end{table}

We compared our extracted SD+compensation spectrum to the state-of-the-art BT-SETTL10 library\footnote{\url{http://phoenix.ens-lyon.fr/Grids/BT-Settl/AGSS2009/SPECTRA/}} \citep{allard2010} following the least-square minimization method of \citet{mohanty2007}. We restrained the fit to 500~K $\leqslant \mathrm{T}_{e\!f\!f} \leqslant$ 1500~K and 3.5 $\leqslant \mathrm{log}\:g \leqslant$ 5.5, assuming metallicity [M/H]~=~0~dex. The best-fit solutions were checked visually. Figure~\ref{fig:compare_spectrum} presents the comparison of our SD+compensation spectrum to synthetic spectra covering \teff from 700~K to 1300~K and \logg from 4.5 to 5.5~dex, and Table~\ref{tab:spectral_analysis} (lines labeled ``SD+compensation'') gives the different estimations of \teff and \logg for our SD+compensation spectrum. We also calculated the H$_{2}$O and CH$_{4}$ spectral indices defined by \citet{burgasser2006} for the classification of T dwarfs and present our results in Fig.~\ref{fig:spectral_indexes}.

In $H$-band, the 1.48--1.57~\mic and 1.72--1.80~\mic regions of the companion spectrum are simultaneously reproduced by models at \teff~=~900--1100~K. The band-head shape between 1.48 and 1.62~\mic is most closely matched at \teff~=~900~K. The models however fail to reproduce the strength and the details of the absorption from 1.62~\mic to 1.73~\mic. This mismatch is also seen in Fig.~23 of \citet{king2010} for $\epsilon$~Indi~Ba/Bb and is interpreted as the result of incomplete CH$_{4}$ line lists. The mismatch is less significant at high surface gravities (\logg~=~5.0--5.5). As a consequence, we present only results when this region of the spectrum was ignored in the comparison to the BT-SETTL10 library.

The analysis of the $J$-band spectrum is more tentative given the uncertainties associated with the extraction (Sect.~\ref{sec:validation_method}). The models notably confirm a lack of flux between 1.0~\mic and 1.18~\mic in the companion spectrum. We note however that the shape of the spectrum between 1.0~\mic and 1.1~\mic can only be reproduced by models with \teff~$\ge$~800~K and \logg~$\ge$~4.5. We then finally estimate \teff~=~$1000 \pm 100$~K and \logg~$\ge$~4.5 for SCR~1845~B.

The spectral indices calculated in our spectrum (Fig.~\ref{fig:spectral_indexes}, brown line) closely agree with the previously assumed T6 spectral type. The H$_{2}$O-$J$, H$_{2}$O-$H$, and CH$_{4}$-$H$ indices yield a T6~$\pm$~1 spectral type. The CH$_{4}$-$J$ index could not be calculated because of the strong over-subtraction in 1.31--1.34~\mic that prevented us from recovering a usable spectrum. However, for that spectral index, our alternative method, which did not show a flux depletion as large as the SD spectrum, yielded a T4--T5 spectral type that agrees with the other indices.

The spectroscopic \teff agrees well with the value of \teff~=~$1042 \pm 162$~K found using the optical/NIR conversion scale of \citet{stephens2009} and assuming a T6~$\pm$~1 spectral type. Our atmospheric parameters are also consistent with those found for field dwarfs by \citet{testi2009} for this spectral type range (1000--1400~K), and based on AMES-COND modelx fitting \citep{allard2001}. Finally, they also agree within the error bars\footnote{Note that \citet{kasper2007} do not state clear error bars for \teff and \logg in their work, so our results agree \emph{within our error bars}.} with the values reported by \citet{kasper2007}.

\begin{table}
  \caption{Summary of the results obtained with the different spectra.}
  \label{tab:summary_results}
  \centering
  \begin{tabular}{lccl}
  \hline\hline
  \multicolumn{1}{c}{Spectrum} &     \teff &    \logg & \multicolumn{1}{c}{SpT} \\ 
                      &       (K) &    (cgs) &      \\
  \hline
  SD+compensation     & 900--1100 & 4.0--5.5 & T5.0--6.5 \\
  No star subtraction &      1200 & 4.0--5.5 & T2.5--4.5 \\
  Alt. method         & 900--1100 & 4.0--5.0 & T4.0--5.5 \\
  \citet{kasper2007}  & 900--1100 & 4.5--5.5 & T5.0--6.0 \\
  \hline
  \end{tabular}
\end{table}

\subsection{Luminosity, radius, and mass}
\label{sec:luminosity_radius_mass}

We found a luminosity $\log(L/L_{\odot}) = -5.28 \pm 0.08$~dex using the parallax of \citet{henry2006}, $BC_{K}$ of \citet{golimowski2004} for a T6~$\pm$~1 ($2.20 \pm 0.12$~mag), the $K$-band magnitude from \citet{kasper2007}, and their associated errors. The values of \teff and luminosity were combined to derive semi-empirical radii using the Stefan-Boltzmann formulae $L = 4\pi R^{2}\sigma \mathrm{T}_{e\!f\!f}^{4}$. With \teff~=~$1000 \pm 100$~K and $\log(L/L_{\odot}) = -5.28 \pm 0.08$~dex, we estimate a radius R~=~$0.7 \pm 0.1$~\RJup. We find a similar value (R~=~$0.7^{+0.2}_{-0.1}$~\RJup) considering \teff~=~$1042 \pm 162$~K. Evolutionary models predictions of \citet{baraffe2003} are consistent with our luminosity, \teff, and \logg estimates for ages $\geq 0.8$~Gyr. The predicted and the semi-empirical radii agree for ages  $\geq 1.5$~Gyr. These constraints on the system age are fully consistent with the one derived by \citet{kasper2007}. Assuming an age between 1.5~Gyr and 10~Gyr, our values of \teff and luminosities correspond to masses between 25 and 65~\MJup. This mass range  also corresponds to the estimates made by \citet{kasper2007} using \citet{burrows1997} evolutionary tracks.

\section{Discussion}
\label{sec:discussion}

Throughout this work, we have demonstrated that the SD method is efficient at suppressing the stellar light contribution in high-contrast spectra. We now discuss the impact of this suppression on the determination of the spectral type and the spectral analysis. We also discuss the advantages of SD compared to other methods that have been used in different contexts.

\subsection{Impact of suppressing the stellar contribution}
\label{sec:impact_suppressing_stellar_contribution}

As we have seen at the end of Sect.~\ref{sec:validation_method}, the SD+compensation spectrum does not match the spectrum extracted without subtracting the star. We quantified it mathematically using the $\epsilon_{\mathrm{rel}}$ factor (Eq.~\ref{eq:relative_error}), but it is also important to quantify the impact of a biased spectrum on the spectral analysis. For this purpose, we performed the spectral analysis (Table~\ref{tab:spectral_analysis}) and calculated spectral indices (Fig.~\ref{fig:spectral_indexes}) in the different spectra that were available: the SD+compensation spectrum, the SD-only spectrum, the spectrum without subtracting the star, the spectrum obtained with our alternative method, and finally the spectrum of \citet{kasper2007}. The only exception is that spectral indices were not calculated for the SD-only spectrum because of the unphysical negative values that would produce spurious values.

The \teff (900--1100~K) and \logg (4.5--5.5) estimated from the SD+compensation and the alternative method are in good agreement with the \citet{kasper2007} results and with the new fits we obtained from their data (bottom of Table~\ref{tab:spectral_analysis}). However, their $J$-band data provides a closer fit to the BT-SETTL10 models than our $J$-band data below 1.18~\mic, yielding a value of \logg that is slightly higher (4.5), which is more in agreement with the $H$-band data. The H$_{2}$O-$J$, H$_{2}$O-$H$, and CH$_{4}$-$H$ spectral indexes are also in excellent agreement (differing by at most 0.05). We can then conclude that the overall agreement between our SD+compensation spectrum and previously published results is good for the estimation of the physical parameters.

In contrast, the spectral analysis performed on the SD-only spectrum and the spectrum without stellar light subtraction shows a strong bias. The former yields a very low \teff and \logg (600~K, 3.5~dex) in $J$-band and reasonable values in $H$-band, while the latter is clearly biased towards higher \teff in both the $J$- and $H$-band spectra. For the spectral indices, the spectrum with no star subtraction systematically leads to spectral types lower than T6 (T2--T6). In particular, the H$_{2}$O-$J$ index is clearly incompatible with the SD+compensation, alternative method, and \citet{kasper2007} spectra. 

Finally, in Table~\ref{tab:summary_results} we have reported the estimations of \teff, \logg, and spectral types (derived from the spectral indices) with the different spectra. The spectrum without star subtraction is clearly completely biased, while the SD+compensation spectrum is in good agreement with the alternative method spectrum and the \citet{kasper2007} data.

The conclusion of this comparison is that in high-contrast spectroscopic data it is essential to fight two major effects: the contamination of the companion spectrum by the star, and the systematic flux losses caused either by the slit losses or the \emph{a posteriori} data analysis. Both these effects clearly introduce a significant bias in the determination of the spectral type and the physical parameters. 

\subsection{Comparison of SD to other methods}
\label{sec:comparison_sd_other_methods}

In Sect.~\ref{sec:context}, we presented a brief summary of the data analysis methods that have been used to differentiate point sources spectra from a constant or varying background, specifically for high-contrast spectroscopy. Given the data they have been applied to, these methods are perfectly valid for subtracting the contribution of the star at the position of the companion, and they would probably provide identical results to our data for SCR~1845~b. Our alternative data analysis method is indeed inspired by these methods and produces an output similar to previously published results. The SD data analysis method is then perfectly comparable to these other methods in terms of the quality and fidelity of the spectral extraction. However, the aforementioned methods all rely on the same assumptions: the stellar halo is symmetrical with respect to the star, and it can be accurately modeled using simple mathematical functions (polynomials, Gaussian or Moffat profiles, ...). While these assumptions may be reasonable for the current state of high-contrast instrumentation, this will certainly not be the case for upcoming extremely high-contrast spectro-imagers.

In very high-contrast observations, with extreme AO systems and state-of-the-art coronagraphs, the data will be limited by the presence of speckles induced by instrumental aberrations. This will be the main limitation of any type of data (imaging or spectroscopy), in particular because of the temporal variance in the speckle pattern \citep[e.g.][]{hinkley2007}. In LSS data, the speckles will modulate the shape of the star halo by creating oblique lines of varying intensity along the spectral dimension \citep{sparks2002}. This is a major limitation because the signal of planetary companions will likely be at the level or below that of the speckles. In this kind of data, simple data analysis methods will not work: although the overall shape of the star halo is symmetrical and can be modeled with simple functions, the speckles superimposed over the halo are asymmetrical and cannot easily be modeled. 

In this context, the SD is the method of choice to remove the star contribution \citep{vigan2008}, be it halo or speckles. In very high-contrast observations, using fake companions as in Sect.~\ref{sec:impact_over_subtraction} to measure the over-subtraction is impossible because the signal of the true companion is too faint to be measured accurately. However, the over-subtraction should be much less significant for several reasons. The main one is that extreme AO systems will provide truly diffraction-limited images in the near-IR (Strehl ratio $\ge$ 80\%), hence the companion PSF will be fully resolved within the slit. For bright companions, the central peak of the companion PSF (which contains $\sim$80\% of the total energy) will then be much easier to mask in the data analysis process and will not introduce any significant bias. The second reason is that in SPHERE/IRDIS, the slit will be oriented along the star-companion axis. This has an impact on the estimation of the model spectrum that is fitted in each spatial channel to estimate the contribution of the star. In a configuration where the star is kept within the slit, the diffraction pattern and speckles from the star remain within the slit when wavelength increases, so they have the same contribution at all wavelengths. In the configuration that was adopted with NACO, these star-related artifacts cross the slit because this slit is tilted with respect to the star-companion axis, hence their contribution varies much more with wavelength: a speckle that is located within the slit at wavelength $\lambda$ will move out of the slit at wavelength $\lambda+\Delta\lambda$. This effect is small in our data because the variation of the halo is small with wavelength, but the presence of bright speckles would bias the result even more.

\section{Conclusion}
\label{sec:conclusion}

The spectral characterization of sub-stellar companions is particularly important for the understanding of their composition and formation processes. The large contrast ratio of these objects with their parent star is usually a limitation of our ability to extract a clean spectrum, free from any significant contribution of the star. In the context of the future high-contrast specro-imager SPHERE/IRDIS, we have developed a data analysis method designed to estimate and remove the contribution of the stellar halo and speckles. In the present work, we have been able to demonstrate the use of this method with VLT/NACO data of SCR~1845~B. 

We have shown that the SD method allows proper recovery of the overall spectral slope, as well as the depth of the spectral features. When good conditions are met (proper sampling, good observing conditions), it provides a result identical to an independent extraction method and very close to the spectrum of SCR~1845~B previously published. We demonstrated that the spectral analysis is strongly biased if the stellar contribution is not removed, asserting the essential need for the data analysis step. However, the over-subtraction of the companion signal in our data makes it necessary to use a fake companion spectrum to estimate and compensate for the loss of flux. This effect is partly due to the companion PSF being larger than the slit, particularly so at lower wavelengths where the AO correction is less efficient.

From our SD+compensation spectrum, we infer a spectral type of T6~$\pm$~1, which agrees well with previous studies and we estimate \teff~$= 1000 \pm 100$~K and \logg~=~4.5--5.5~dex by comparing our data to the state-of-the-art BT-SETTL10 library of models. With a luminosity $\log(L/L_{\odot}) = -5.28 \pm 0.08$~dex, we estimate a radius R~=~$0.7^{+0.2}_{-0.1}$~\RJup, which is consistent with evolutionary models for ages older than 1.5~Gyr 

Finally, these results represent an important milestone for the use of the SD method at very high-contrast for the spectral characterization of planetary-mass objects with LSS and coronagraphy in VLT/IRDIS. Data of future high-contrast imagers and spectrographs will indeed be modulated by the presence of instrumental speckles that cannot be estimated and removed using simple methods, increasing the need for more sophisticated data analysis methods.

\begin{acknowledgements}
A. Vigan acknowledges support from a Science and Technology Facilities Council (STFC) grant (ST/H002707/1). We are grateful to M. Kasper for providing his published spectrum, to F. Allard for providing us with the SETTL10 library of models, to A. Burgasser for maintaining the SpecX Prism libary, and finally to the ESO staff for performing the service mode observations.
\end{acknowledgements}

\bibliographystyle{aa}
\bibliography{paper}

\end{document}